\documentclass[prb,showpacs,twocolumn,preprintnumbers,superscriptaddress]{revtex4}
\usepackage{graphicx}
\usepackage{bm}

\def\beq{\begin{equation}}
\def\eeq{\end{equation}}
\def\beqn{\begin{eqnarray}}
\def\eeqn{\end{eqnarray}}

\newcommand{\nv}[1]{{\bm #1}}
\newcommand{\up}{\uparrow}
\newcommand{\down}{\downarrow}

\begin{document}

\title{
Variational cluster approach to spontaneous symmetry breaking: \\
The itinerant antiferromagnet in two dimensions
}

\author{C. Dahnken}
\affiliation{Institut f\"ur Theoretische Physik und Astrophysik,
Universit\"at W\"urzburg, Am Hubland, D-97074 W\"urzburg, Germany}

\author{M. Aichhorn}
\affiliation{Institut f\"ur Theoretische Physik,
Technische Universit\"at Graz, Petersgasse 16, A-8010 Graz, Austria}

\author{W. Hanke}
\affiliation{Institut f\"ur Theoretische Physik und Astrophysik,
Universit\"at W\"urzburg, Am Hubland, D-97074 W\"urzburg, Germany}

\author{E. Arrigoni}
\affiliation{Institut f\"ur Theoretische Physik,
Technische Universit\"at Graz, Petersgasse 16, A-8010 Graz, Austria}

\author{M. Potthoff}
\email{potthoff@physik.uni-wuerzburg.de}
\affiliation{Institut f\"ur Theoretische Physik und Astrophysik,
Universit\"at W\"urzburg, Am Hubland, D-97074 W\"urzburg, Germany}

\date{\today}

\begin{abstract}
Based on the self-energy-functional approach proposed recently
[M. Potthoff, Eur. Phys. J. B {\bf 32}, 429 (2003)], we present 
an extension of the cluster-perturbation theory to systems with 
spontaneously broken symmetry.
Our method 
applies to models with local interactions and 
accounts for both short-range correlations and 
long-range order.
Short-range correlations are accurately taken into account via
exact diagonalization of finite clusters.
Long-range order is described by variational optimization of
a ficticious symmetry-breaking field.
In comparison with related cluster methods,
our approach is more flexible and, for a given cluster size, less 
demanding numerically, especially at zero temperature.
An application of the method to the antiferromagnetic phase of
the Hubbard model at half-filling shows good agreement with 
results from quantum Monte-Carlo calculations.
We demonstrate that the variational extension of the cluster-perturbation 
theory is crucial to reproduce salient features of the single-particle 
spectrum.
\end{abstract}


\maketitle

\section{Introduction}

Several materials studied in condensed-matter physics display unusual physical
properties which are ascribed to strong electron correlations.
\cite{OM00,IFT98,BDN01}
In particular, these may give rise to rich phase diagrams with different
competing types of short-range correlations and with different 
symmetry-broken phases.
Realistic many-body models of these materials contain an interaction-energy 
term of the same order of magnitude as the kinetic energy or even larger.
This implies that it is quite generally inappropriate to treat these models 
by conventional weak-coupling perturbation theory or by static mean-field 
decouplings such as the Hartree-Fock approximation.

A complementary approach consists in an exact treatment of the interacting 
part while the kinetic energy is accounted for perturbatively.
For Hubbard-type models, this corresponds to an expansion in powers of 
the hopping $t$ around the atomic limit, and leads - at the lowest 
order - to the so-called Hubbard-I approximation. \cite{hubb.63}
An expansion in $t$ can be organized in a systematic diagrammatic 
series. \cite{metz.91,pa.se.98,pa.se.00}
This approach, however, not only fails for $t$ of the order of the
Hubbard repulsion $U$, but also for low temperatures, due to the 
degeneracy of the ground state.

An interesting extension of this strong-coupling
expansion consists in dividing the lattice into clusters of 
sufficiently small size such that they can be treated exactly, 
followed by an expansion in powers of the hopping between the 
clusters. \cite{gr.va.93,se.pe.00,se.pe.02}
The expansion in the inter-cluster hopping can be formally 
carried out up to arbitrary order following the diagrammatic method 
of Refs.~\onlinecite{metz.91,pa.se.98,pa.se.00}.
However, going beyond the lowest order, is quite demanding
numerically and has not been carried out so far for two-dimensional
systems. 
For one-dimensional (infinitely long) chains, on the other hand, 
such an expansion could be accomplished in fact -- to infinite order 
in powers of the inter-chain 
hopping. \cite{arri}

The lowest order of the strong-coupling expansion in the inter-cluster
hopping \cite{gr.va.93} has been termed ``cluster-perturbation theory'' 
(CPT). \cite{se.pe.00}
Actually, the CPT can be considered as a systematic approach with respect 
to the cluster size, i.e.\ it becomes exact in the limit 
$N_{\rm c} \to \infty$ where $N_{\rm c}$ is the number of sites within
a cluster.
From this point of view the CPT represents an attractive method which is 
simple conceptually but nevertheless includes short-range correlations 
on the scale of the cluster size.
Moreover, the CPT provides results for an infinitely extended system.
Consequently, the CPT Green's function is defined for any wave vector 
${\bm k}$ in the Brillouin zone, contrary to common ``direct'' cluster 
calculations for which only a few ${\bm k}$ points are available, 
The necessary numerical effort is moderate:
Once the Green's function of a cluster of a given size has been calculated
via a numerical method, e.g.\ the Lanczos technique, the determination of the 
lattice Green's function is numerically much less demanding as this requires 
the inversion of a certain number of matrices with a dimension given by 
$N_{\rm c}$ only.

CPT results for static quantities as well as for the single-particle
spectral function have been shown to agree well with different exact 
analytical and numerical results. \cite{se.pe.00,se.pe.02}
Recently, a generalization of the method with different cluster shapes has
successfully been used for an analysis of the stripe phase in 
high-temperature
superconductors. \cite{za.ed}
On the other hand, there is also a serious disadvantage of the CPT at this 
level: Namely, the method does not contain any self-consistent procedure
which implies that symmetry-broken phases cannot be studied (the case of a 
degenerate ground state, as e.g.\ for the Hubbard-I approximation, 
represents an exception).

This generates the motivation for the present paper.
We will present an extension of the CPT which is based on the 
self-energy-functional approach (SFA) proposed recently. \cite{Pot03a}
The SFA provides a general variational scheme to use dynamical 
information from an exactly solvable reference system (an isolated cluster)
to approximate the physics of a system in the thermodynamic limit.
Using the SFA it is possible to construct a self-consistent or variational 
cluster-perturbation theory (``V-CPT'') which allows to study phases with 
spontaneously broken symmetry.
The V-CPT applies to arbitrary Hubbard-type lattice models with the restriction 
that the interaction be local.

Self-consistent cluster methods can also be constructed as generalizations of
the dynamical mean-field theory (DMFT) \cite{me.vo.89,ge.ko.96} as has been shown
in recent years. \cite{SI95,HTZ+98,LK00,ko.sa.01,ma.ja.02,bi.ko.02}
Similar as the CPT, the cellular dynamical mean-field theory (C-DMFT) 
\cite{ko.sa.01} is based on a real-space formulation. 
The C-DMFT performs a self-consistent mapping of the lattice problem onto an 
effective cluster model with $N_{\rm c} > 1$ correlated sites and reduces to the 
standard DMFT for $N_{\rm c} = 1$. 
CPT and C-DMFT differ with respect to the concept of ``bath'' sites.
The effective cluster model which is considered in the C-DMFT contains an
infinite number of additional uncorrelated (``bath'') sites attached to each 
of the $N_{\rm c}$ original correlated sites in the cluster.
The bath parameters are determined from a self-consistency condition.
This construction ensures an optimal description of the local (temporal)
degrees of freedom but complicates the method considerably.
A numerically exact evaluation of cluster generalizations of the DMFT could
so far only be achieved by using quantum Monte-Carlo techniques.
\cite{JMHM01,MPJ02}

Recently, it has been pointed out \cite{PAD03} within the context of the 
self-energy-functional approach (SFA) that both a (variational) CPT and 
the C-DMFT can be considered as extreme limits ($n_{\rm s}=1$ and 
$n_{\rm s}=\infty$) of a more general cluster method where reference is 
made to an effective cluster model with $N_{\rm c}$ correlated sites and 
$n_{\rm s} - 1$ additional bath sites 
per correlated site.
Hence, the SFA formally unifies the different cluster approaches and thereby 
places our proposed method in a more general context.
SFA-cluster calculations for the one-dimensional Hubbard model \cite{PAD03} 
strongly suggest that it is more efficient to use a cluster as large as possible 
and set $n_{\rm s}=1$ (no bath sites) -- as compared to a smaller cluster and $n_{\rm s} > 1$.
This is contrary to the opposite limit of infinite dimensions:
For $D=\infty$ the exact theory (namely DMFT) is obtained for $N_{\rm c}=1$ and 
$n_{\rm s}=\infty$.

It is, therefore, particularly interesting to apply the V-CPT to the 
two-dimensional case and to compare with available numerically exact results.
The low-temperature antiferromagnetic phase of the $D=2$ Hubbard model at 
half-filling represents an optimal playground to study the strengths and 
limitations of the method. The reason is that both the effects of short-range 
correlations and long-range antiferromagnetic order manifest themselves 
in static thermodynamic quantities as well as in the single-particle excitation
spectrum.

The paper is organized as follows: 
The variational generalization of the CPT is introduced in the following 
section.
In Sec.~\ref{resu} we present our results for the antiferromagnetic $D=2$ 
Hubbard model. 
One- and quasi-one-dimensional systems are discussed briefly, in addition.
The performance of the method is analyzed by comparing with numerical 
results from different methods.
Emphasis is given to the single-particle excitation spectrum.
Finally, our conclusions and an outlook are presented in Sec.~\ref{conc}.

\section{Variational CPT}
\label{meth}

Consider a system of interacting electrons on a lattice with a Hamiltonian $H$ 
consisting of a single-particle (non-interacting) term $H_0$ and an interaction 
term $H_1$. 
We require that the interacting part be local.
This allows for a partitioning of the lattice into non-overlapping clusters of
finite size which are not connected by $H_1$.
In the simplest case, $H_1$ describes an on-site Hubbard repulsion.
After having divided the lattice into clusters (labeled by $\nv R$), the 
Hamiltonian can be written in the form
\beq
\label{h}
 H = \sum_{\nv R} \left[
 H^{\rm (intra)}_0(\nv R) + H_1(\nv R) \right]
 + \sum_{\nv R,\nv R'}^{\nv R \ne \nv R'} H_0^{\rm (inter)}(\nv R,\nv R') \; ,
\eeq
where
\beq
\label{h0}
H_0^{\rm (intra)}(\nv R) = \sum_{a,b} t_{a,b} \: c^{\dag}_{\nv R a}
c_{\nv R b} 
\eeq
is the non-interacting, intra-cluster part of the Hamiltonian, and $H_1(\nv R)$
is the intra-cluster interaction part (which we do not need to specify). 
The remaining term
\beq
\label{t}
H^{\rm (inter)}_0(\nv R,\nv R') = \sum_{a,b} V_{\nv R a, \nv R'b} \:
c^{\dag}_{\nv R a} c_{\nv R' b} \;,
\eeq
is a non-interacting part connecting different clusters (inter-cluster hopping).
The labels $a,b$ indicate positions within a cluster as well as other (spin and 
orbital) degrees of freedom.
$c_{\nv R a}$ annihilates an electron with quantum numbers $a$ within the cluster 
$\nv R$.
For simplicity, translational invariance with respect to the ``superlattice''
vector $\nv R$ is assumed.

We are interested in the single-particle Green's function 
$G_{\nv R a, \nv R'b}(\omega) = 
\langle \langle c_{\nv R a} ; c^{\dag}_{\nv R' b} \rangle \rangle_{\omega}$.
Exploiting translational invariance and performing a Fourier transformation
to the reciprocal space, the Green's function becomes diagonal with respect to
the wave vector $\nv Q$ from the (reduced) Brillouin zone corresponding to 
the superlattice. 
In reciprocal space, the Green's function is a matrix $\nv G_{\nv Q}(\omega)$ 
with elements $G_{\nv Q, a, b}(\omega)$ labeled by the cluster variables 
$a$ and $b$.

Let us define a ``reference system'' with Hamiltonian $H'$ where the inter-cluster
hopping $H^{\rm (inter)}_0$ is switched off:
\beq
  H' = \sum_{\nv R} \left[
  H^{\rm (intra)}_0(\nv R) + H_1(\nv R) \right] \; .
\eeq
$H'$ describes a system of decoupled clusters of finite size.
For not too large clusters, this system can be solved exactly, and its 
Green's function $\nv G'(\omega)$ can be computed by conventional 
methods such as exact diagonalization (ED) or quantum Monte-Carlo (QMC).
Generally, the corresponding Green's function $\nv G'(\omega)$ is a matrix 
with indices $(\nv R a)$ and $(\nv R' b)$. 
As $H^{\rm (inter)}_0=0$ this matrix is diagonal (and constant) with respect
to $\nv R$:
\beq
  \langle \langle c_{\nv R a} ; c^{\dag}_{\nv R' b} \rangle \rangle'_{\omega} = 
  \delta_{\nv R,\nv R'} \; G'_{a,b}(\omega)  \; .
\eeq

Within the CPT approximation, the Green's function $\nv G(\omega)$ of the full 
problem $H$ is expressed in terms of $\nv G'(\omega)$ and the inter-cluster hopping 
$V_{\nv R a, \nv R'b}$ by an RPA-type expression: \cite{gr.va.93,se.pe.00}
\beq
\label{gcpt}
\nv G_{\nv Q}(\omega) = \left[ \nv G'(\omega)^{-1} - \nv V_{\nv Q}
\right]^{-1} \; .
\eeq
Here, $\nv G_{\nv Q}(\omega)$, $\nv G'(\omega)$, and $\nv V_{\nv Q}$ 
are matrices in the cluster indices $a$ and $b$.
The Fourier-transformed inter-cluster hopping is given by:
\beq
\label{vq}
  V_{\nv Q, a,b} = \frac{1}{L} \sum_{\nv R, \nv R'} V_{\nv R a, \nv R' b} \;
  e^{i \nv Q \cdot (\nv R - \nv R')} \; ,
\eeq
where $L$ is the number of superlattice sites.

The above formalism constitutes the ``usual'' CPT approach. 
We like to stress that the method is based on the exact solution of 
finite-size clusters in which spontaneous symmetry breaking cannot occur.
Furthermore, it does not include any self-consistent procedure.
Consequently, symmetry-broken phases cannot be studied within the usual CPT.

Our proposal for a proper generalization of the CPT is the following:
First, one should note that in the CPT the perturbation term is quite arbitrary 
and can be taken as {\em any} one-particle operator. 
The partition of the non-interacting part of the Hamiltonian Eq.\ (\ref{h}) 
gives 
us a certain amount of freedom that we can exploit to seek for an optimized 
starting point.
As a matter of fact, one has the freedom to add to $H_0^{\rm (intra)}$ any
local single-particle term which is then subtracted in $H_0^{\rm (inter)}$.
In other words, the Hamiltonian (\ref{h}) is obviously invariant under the
transformation
\beqn
\label{tras}
H^{\rm (intra)}_0(\nv R) & \to & H_0^{\rm (intra)}(\nv R) + \Delta(\nv R) 
\nonumber \\ 
H^{\rm (inter)}_0(\nv R, \nv R') & \to & H^{\rm (inter)}_0(\nv R,\nv R') 
- \delta_{\nv R,\nv R'} \ \Delta(\nv R) \; ,
\nonumber \\ 
\eeqn
where $\Delta(\nv R)$ is an arbitrary intra-cluster single-particle operator
which can be expressed as
\beq
\label{delta}
\Delta(\nv R) = \sum_{a,b} \Delta_{a,b} \ c^{\dag}_{\nv R a} c_{\nv  R b} \;.
\eeq
Formally, the rest of the procedure remains unchanged.

If the perturbative approach was exact, the results would not depend on
$\nv \Delta$ at all.
As a matter of fact, this can easily be seen in the non-interacting limit 
in which the CPT becomes exact.
\footnote{
This is due to the fact that, as in the expansion around the atomic limit 
of the Hubbard model, corrections beyond the CPT contain cumulants
of higher-order correlation functions that vanish in the
noninteracting limit. \cite{metz.91}
}
Here the perturbation cancels out.

In the interacting case, the result {\it does} depend on $\nv \Delta$.
However, this is not a shortcoming. On the contrary, this allows
us to ``optimize'' the results of the CPT calculation.
\footnote{
This approach has been recently suggested to introduce periodic
boundary conditions. \cite{DAH02}
Hopping terms connecting the cluster boundaries are added to the 
Hamiltonian for the isolated cluster and then subtracted again within
the CPT scheme. 
This procedure has turned out not to be satisfactory, however.
Indeed it has been shown in Ref.\ \onlinecite{PAD03}
that open boundary conditions should be used for the variational CPT.
}
Indeed, we may think of choosing $\nv \Delta$ such that the single-particle
dynamics of the cluster problem is ``as close as possible'' to the
exact dynamics of the lattice. In this way, one can hope that the
perturbative correction is small and that the result of the
perturbative calculation is accurate enough.
The question is how to perform this optimization in practice.
Note that the answer to this question also solves our original problem as one 
may choose the perturbation $\nv \Delta$ to represent a ``ficticious'' 
symmetry-breaking field term (a staggered magnetic field, for example)
since this has the form of a one-particle operator.

A straightforward idea to optimize $\nv \Delta$ (or the strength $h$ of the
symmetry-breaking field) would be to express a thermodynamical potential 
(the grand potential $\Omega$, for example) in terms of the CPT Green's function
${\bm G}$, which depends on $\nv \Delta$ in turn, and to minimize the function
$\Omega[{\bm G}(\nv \Delta)]$ with respect to $\nv \Delta$. 
However, the following serious problems arise:
As the CPT Green's function is approximate and as there are different ways to 
obtain the grand potential, the procedure is not unique.
So there are several ways the potential can depend on the perturbation, and the 
results will depend on the respective choice.
Moreover, once the grand potential is given in terms of $\nv \Delta$, 
there is no physical reason to minimize the grand potential as this would 
require a corresponding variational principle to be valid which is generally 
not the case. 
Below, however, we will show that an appropriate variational principle can be found 
in fact, and a corresponding potential, Eq.\ (\ref{omega}), can be constructed the stationary 
point of which gives an optimized $\nv \Delta$.

Exact variational principles of the form $\delta \Omega[{\bm G}] = 0$ or 
$\delta \Omega[{\bm \Sigma}] = 0$ where ${\bm \Sigma}$ is the self-energy 
are actually known for a long time from standard diagrammatic theory.
\cite{lu.wa.60}
The problem is that the functional dependence $\Omega[{\bm G}]$ or 
$\Omega[{\bm \Sigma}]$ is not given explicitly but has to be constructed via
an infinite sum of renormalized skeleton diagrams.
This has impeded the use of the variational principles in their original form.

Here, the help comes from the self-energy-functional approach (SFA) \cite{Pot03a}
proposed recently.
The SFA provides a way to exactly evaluate the functional $\Omega[{\bm \Sigma}]$ 
-- even if the functional dependence is not explicit.
This is achieved at the cost of a restriction of the domain of the functional,
i.e.\ the functional $\Omega[{\bm \Sigma}]$ can be evaluated exactly on
a certain subspace ${\cal S}$ of trial self-energies.
The idea is then to perform a search for the stationary point on the restricted 
space ${\cal S}$.
The subspace ${\cal S}$ consists of all $\nv \Sigma$ which are exact self-energies
of a reference system.
Clearly, the Hamiltonian of the reference system $H'$ must be exactly solvable 
so that one is able to compute the self-energy in practice.
Furthermore, general arguments \cite{Pot03a} require that $H'$ must have the 
same interaction part as $H$. 
The one-particle part of the reference system, however, is completely arbitrary 
and its parameters may be used to optimize the trial self-energy.
Note that these conditions are fulfilled for the case considered here.
Constructing the reference system by dividing the lattice into small clusters,
the SFA concept just yields the desired variational CPT.

To verify this, it is sufficient to realize the following:
Equation (\ref{gcpt}) which approximates the Green's function $\nv G_{\nv Q}(\omega)$
of $H$ in terms of the Green's function $\nv G'(\omega)$ of the system of decoupled 
clusters $H'$ and the inter-cluster hopping $\nv V_{\nv Q}$, can be cast into the 
form of a Dyson equation, 
\beq
\label{gcpteq}
\nv G_{\nv Q}(\omega) = (\nv G^{(0)}_{\nv Q}(\omega)^{-1} - 
{\bm \Sigma(\omega)})^{-1} \;.
\eeq
Here $\nv G^{(0)}_{\nv Q}(\omega) = (\omega + \mu - \nv t - \nv V_{\nv Q})^{-1}$ 
is the free Green's function of the infinite lattice given in terms of the
chemical potential $\mu$, and
the intra- and inter-cluster
hopping $\nv t$ and $\nv V_{\nv Q}$, respectively, 
and $\nv \Sigma(\omega)$ is the cluster self-energy.
One can therefore state that the CPT consists in approximating the self-energy 
of the lattice problem $H$ by the ($\nv Q$ independent) self-energy 
${\bm \Sigma(\omega)}$ of the reference system $H'$.

The optimization problem mentioned above is now solved in the following 
way:
Varying $\nv \Delta$ corresponds to varying the one-particle parameters of the 
reference system $H'$, the interaction part being kept fixed.
For any $\nv \Delta$, the reference system can be solved to get the
self-energy $\nv \Sigma$.
Thus, the self-energy is parameterized as $\nv \Sigma = \nv \Sigma(\nv \Delta)$.
Furthermore, the Green's function $\nv G' = \nv G'(\nv \Delta)$ and the grand
potential $\Omega' = \Omega'(\nv \Delta)$ of the reference system $H'$ can
be calculated.  
Following Ref.\ \onlinecite{Pot03a}, the self-energy-functional for the trial 
self-energy $\nv \Sigma(\nv \Delta)$ can be evaluated exactly, i.e.\ 
$\Omega[\nv \Sigma(\nv \Delta)]$ can be calculated. 
This yields a function $\Omega(\nv \Delta) \equiv \Omega[\nv \Sigma(\nv \Delta)]$
the explicit form of which is taken from Ref.\ \onlinecite{Pot03a}:
\beqn
   \Omega(\nv \Delta) &=& \Omega'(\nv \Delta) 
   \nonumber \\
   & + & T \sum_{n} \sum_{\nv Q} \mbox{tr} \ln 
   \frac{- \nv 1}{\nv G^{(0)}_{\nv Q}(i\omega_n)^{-1} - \nv \Sigma(\nv \Delta,i\omega_n)}
   \nonumber \\
   & - & LT \sum_{n} \mbox{tr} \ln (- \nv G'(\nv \Delta,i\omega_n)) \: .
\label{omega}
\eeqn
Here, the frequency sums run over the discrete Matsubara frequencies $i\omega_n$, 
$L$ is the number of clusters (or, equivalently, the number of $\nv Q$ points),
and bold symbols denote matrices with respect to cluster indices $a$ and $b$.
Note that for the evaluation of the grand potential (\ref{omega}) one needs the
CPT Green's function, Eq.\ (\ref{gcpteq}).
Searching for the stationary point of the function $\Omega(\nv \Delta)$ 
means to search for the
stationary point of the exact self-energy functional on the restricted domain
of $H'$-representable self-energies.
This prescription tells us which approximate cluster self-energy as best as 
possible describes the exact one.

\section{Results}
\label{resu}

We have applied the variational CPT (V-CPT) presented above to the single-band
Hubbard model at half-filling and zero temperature.
The Hamiltonian reads:
\beq
\label{hubb}
H =  \sum_{\nv r,\nv r'} t_{\nv r , \nv r'} \ c^{\dag}_{\nv r,\sigma}
c_{\nv r',\sigma}
+ U \sum_{\nv r} n_{\nv r,\up} n_{\nv r,\down} \: .
\eeq
Here $c_{\nv r,\sigma}$ annihilates an electron with spin projection 
$\sigma=\uparrow,\downarrow$ at
the lattice site $\nv r$, $n_{\nv r,\sigma} = c^{\dag}_{\nv r,\sigma}
c_{\nv r,\sigma}$, and the hopping 
$t_{\nv r,\nv r'}$ is equal to $t$ and nonvanishing for nearest-neighbor 
sites $\langle \nv r$, $\nv r' \rangle$ only.
Throughout the paper, $t=1$ sets the energy scale.

In principle, the ``best'' result is obtained by using a completely general 
single-particle term $\nv \Delta$. However, this would imply the computation of 
$\Omega(\nv \Delta)$ for a too large number of parameters making the problem 
numerically impractical.
For this reason, it is more convenient to start with a ``guess'' of the 
appropriate physical symmetry-breaking field.
For half-filling, a good candidate is certainly a staggered field producing a 
N\'eel ordered state. 
For simplicity, we consider clusters containing an integer number of 
antiferromagnetic unit cells.
With the notation $a=(\nv r,\sigma)$, the corresponding $\nv \Delta$ 
has the form
\beq
\label{deltah}
\Delta_{a,b} = h \ \delta_{a,b} \ z_\sigma \ \eta_{\nv r} \;,
\eeq
where $\eta_{\nv r}=+1$ ($=-1$) on sites of sublattice $A$ ($B$), 
$z_\sigma = \pm 1$ for spin projection $\sigma= \uparrow,\downarrow$, 
and $h$ is the strength of the ficticious staggered field. 
The optimal value of $h$ will be obtained by minimization 
\footnote{
The physical self-energy is given by a stationary point of 
$\Omega[\nv \Sigma]$. This can be a minimum, maximum or a saddle point.
\cite{Pot03a} For a one-dimensional parameterization, $\nv \Sigma =
\nv \Sigma(h)$, there are minima and maxima in general. In case of
several stationary points, the one with lowest $\Omega$ is stable
thermodynamically. In case of a single parameter $h$ this must be
always at a minimum.
}
of $\Omega(h) = \Omega[\nv \Sigma(h)]$ as given by Eq.\ (\ref{omega}).
Obviously, $h=0$ corresponds to the usual CPT approximation.
We stress again that via the transformation (\ref{tras}), the staggered field
is strictly equal to zero in the original Hamiltonian (\ref{h}). 
It only appears in an intermediate step in the Hamiltonian of the reference
system $H'$ to parameterize the trial self-energy.
Thus, $h$ is a variational parameter without a direct physical meaning
in the original lattice Hamiltonian $H$. However, it does introduce a
true staggered field in the reference (cluster) Hamiltonian $H'$.

\begin{figure}[t]
  \includegraphics[width=0.45\columnwidth]{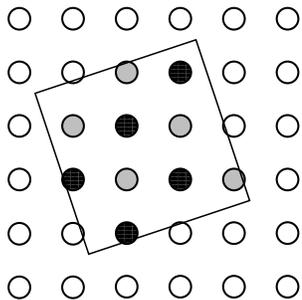}
\caption{
Decomposition of the $D=2$ square lattice into $\sqrt{10}\times\sqrt{10}$ clusters.
}
\label{sqrt10}
\end{figure}

For the numerical calculations we first consider a decomposition of the lattice 
into ``$\sqrt{10}\times\sqrt{10}$'' clusters as indicated in Fig.~\ref{sqrt10}.
Following Ref.\ \onlinecite{PAD03}, open boundary conditions are used.
To evaluate the self-energy functional, the grand potential $\Omega'(h)$ and 
the Green's function ${\bm G}'(h)$ for a cluster are computed using the standard 
Lanczos algorithm. \cite{LG93}
The self-energy is obtained as ${\bm \Sigma}(h) = {{\bm G}'_0}(h)^{-1} - 
{{\bm G}'}(h)^{-1}$.
The sum over Matsubara frequencies in Eq.\ (\ref{omega}) can be transformed
into an integral over real frequencies. \cite{pott.03u.se}
After frequency integration
and $\bm Q$ summation, we obtain $\Omega(h)$ from Eq.\ (\ref{omega}).
A Lorentzian broadening $\omega \to \omega + i \delta$ with finite $\delta = 0.1$
is used.
For this choice typically 500 $\nv Q$ points are sufficient for convergence of
the results.
We have checked that the results do not significantly depend on $\delta$.

\subsection{Static quantities}

\begin{figure}[b]
  \includegraphics[width=0.9\columnwidth]{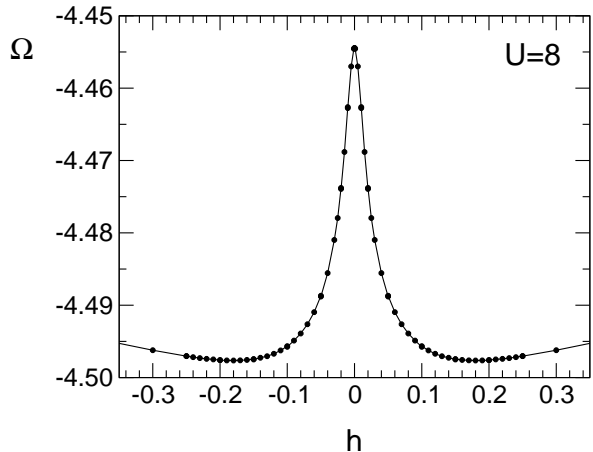}
\caption{
Dependence of the grand potential $\Omega$ (per site)
on the ficticious staggered 
field $h$ as obtained by evaluating the self-energy functional 
$\Omega=\Omega[\nv \Sigma(h)]$.
The lattice is decomposed into ``$\sqrt{10} \times \sqrt{10}$'' clusters
(see Fig.~\ref{sqrt10}).
Parameters: on-site repulsion $U=8$, temperature $T=0$, half-filling.
The optimal staggered field is found to be $h = \pm 0.18$.    
The energy unit is given by the nearest-neighbor hopping.    
}
\label{omegah}
\end{figure}

Using Eq.\ (\ref{omega}), the grand potential $\Omega(h) = \Omega[\nv \Sigma(h)]$
has been calculated for $U=8$ at half filling and $T=0$.
The result is shown in Fig.~\ref{omegah}.
As anticipated, $\Omega$ depends on $h$. Three stationary points are
found: a maximum at $h=0$ and two (equivalent) minima for nonvanishing values
$h \approx \pm 0.18$.
This means that the interacting system ``prefers'' a symmetry-broken state
with a non-vanishing staggered magnetization $m$,
as one would have expected physically.

The stationary point at $h=0$ corresponds to the usual CPT.
For $h=0$ the ground state of a single cluster shows antiferromagnetic 
correlations but is non-degenerate. 
Hence, the cluster Green's function and the self-energy are spin independent.
This implies that there is no {\em coherent} continuation of the antiferromagnetic 
correlations across the cluster boundaries within the usual CPT.
Consequently, the order parameter $m=0$.
The binding energy per site that is gained by a coherent matching of 
antiferromagnetic clusters for finite $h$, can be read off from Fig.\ 
\ref{omegah} to be $\Delta \Omega \approx 0.043$.
This is small as compared to $|J| = 4 t^2 / U = 0.5$ as there are contributions
from bonds connecting different clusters only.

\begin{figure}[t]
    \includegraphics[width=0.85\columnwidth]{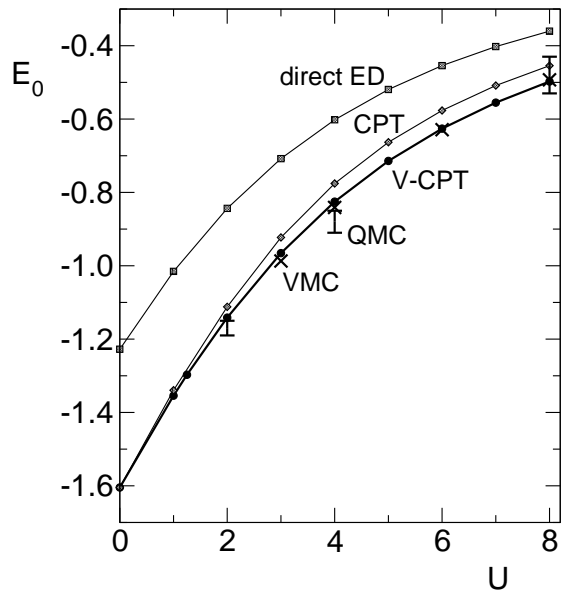}
\caption{$U$ dependence of the ground-state energy per site $E_0$ 
at half-filling and zero temperature as 
obtained by different methods:
direct exact diagonalization (squares), usual (non-variational, $h=0$) CPT 
(diamonds), and variational ($h$ optimized) CPT (V-CPT, circles) 
for $\sqrt{10}\times\sqrt{10}$ clusters.
Additionally, results from a variational Monte Carlo calculation (crosses)
\cite{yo.sh.87b} and a QMC simulation (error bars) 
\cite{hirs.85} are shown for comparison.
VMC and QMC results are extrapolated to $N_{\rm c}=\infty$.
VMC error bars are smaller than the symbol size.
}
\label{gse}
\end{figure}

From the value of the grand potential at the optimal field $h$ the ground-state
energy is obtained as $E_0 = \Omega + \mu \langle N \rangle$.
We have performed calculations for different $U$.
Fig.\ \ref{gse} shows $E_0$ as a function of $U$ for the respective optimal 
ficticious field $h$ (V-CPT) and for $h=0$ (CPT).
The results are compared with those of an exact-diagonalization calculation for
the isolated cluster with $N_{\rm c}=10$ sites (direct ED).
Furthermore, the results of a variational Monte-Carlo (VMC) calculation \cite{yo.sh.87b} 
using a Gutzwiller-projected symmetry-broken trial wave function and the results of
an auxiliary-field quantum Monte-Carlo (QMC) study \cite{hirs.85} are displayed for
comparison.
VMC and QMC data for different cluster sizes $N_{\rm c}$ are extrapolated 
\cite{yo.sh.87b,HT89} to $N_{\rm c} = \infty$ (and to $T=0$, in the
latter case).

As compared to the ground-state energy that is obtained by diagonalization
of an isolated cluster (``direct ED''), the (usual) CPT result 
represents a considerable improvement, as can be seen in the figure.
Note that CPT (and V-CPT) recover the exact result in the non-interacting limit.
The gain in binding energy is due to the (approximate) inclusion of the 
inter-cluster hopping.
A comparison of CPT with Monte-Carlo results (VMC, QMC), however, 
still shows a sizable discrepancy.
On the other hand, our variational CPT method perfectly
agrees within the error bars 
with both Monte-Carlo results for $E_0$ in the entire $U$ range.
This shows that a proper description of long-range order is essential to 
get the ground-state energy accurately.
Note, however, that for the ground state itself and for dynamical quantities, 
the inclusion of short-range correlations is at least equally important
(see below).

\begin{figure}[t]
    \includegraphics[width=0.75\columnwidth]{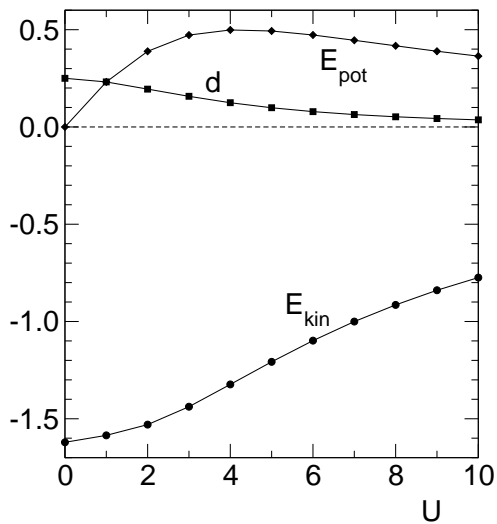}
\caption{
Double occupancy $d=\langle n_{\nv r, \uparrow} n_{\nv r, \downarrow} \rangle$ 
calculated as $d = \partial \Omega(\mu,U) / \partial U$, potential energy per site
$E_{\rm pot} = U \, d$, and kinetic energy per site 
$E_{\rm kin} = E_0 - E_{\rm pot}$ as functions of $U$
obtained for the same system as in Fig.\ \ref{gse} via V-CPT.
}
\label{kdp}
\end{figure}

Let us discuss a few other static quantities.
Fig.\ \ref{kdp} shows the double occupancy 
$d \equiv \langle n_{\nv r\up} n_{\nv r\down} \rangle$ as a function of $U$.
The double occupancy is obtained by numerical differentiation of the 
grand potential $d = \partial \Omega(\mu,U) / \partial U$ (at its
respective minimum value).
It monotonously decreases from the non-interacting value 
$d = \langle n_{\nv r\up} \rangle \langle n_{\nv r\down} \rangle = 0.25$ and 
correctly tends to approach the strong-coupling limit $d=0$.
Already for $U$ of the order of the free band width, a strong suppression 
of $d$ is found ($d \approx 0.052$ for $U=8$).
This indicates a quick crossover from a Slater-type (itinerant moments) to 
a Heisenberg-type antiferromagnet (local moments) with increasing $U$.
The potential energy $E_{\rm pot} = U \, d$ and the kinetic energy 
$E_0 - E_{\rm pot}$ with $E_0 = \Omega + \mu \langle N \rangle$ and 
$\mu = U/2$ are shown in addition.
Despite the fact that local-moment formation is almost completed for $U=8$,
there is still a considerable kinetic energy $E_{\rm kin} \approx -0.915$.
This has to be attributed to the residual kinetic exchange.

\subsection{One-dimensional case}

One may ask whether or not the variational procedure always yields an
antiferromagnetic state, i.e.\ also in those cases in which this is not 
expected physically.
For example, an antiferromagnetic state is prohibited in one dimension as 
quantum fluctuations break up any long-range spin order. \cite{And52}
Mean-field methods, such as Hartree Fock, however, often yield a N\'eel state 
also in one dimension. 
In a strict mean-field theory, spatial correlations are neglected altogether.
Due to the inclusion of short-range correlations, the variational CPT is 
clearly superior as compared to mean-field theory.
For any finite $N_{\rm c}$, however, longer-range spatial correlations are 
neglected.
Hence, the V-CPT may be considered as a mean-field approach on a length scale 
exceeding the cluster dimensions.
We therefore expect ``mild'' reminiscences of typical mean-field artifacts.

\begin{figure}[t]
    \includegraphics[width=\columnwidth]{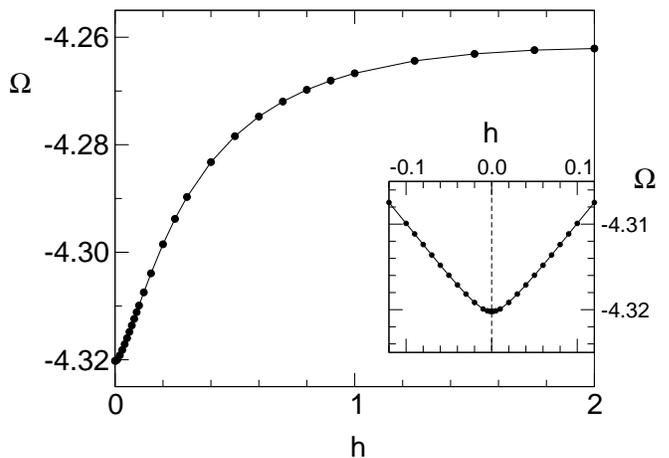}
\caption{Dependence of the grand potential $\Omega$ (per site)
on the ficticious staggered 
field $h$ as obtained by evaluating the self-energy functional 
$\Omega=\Omega[\nv \Sigma(h)]$.
Result for the $D=1$ Hubbard model at $U = 8$, $T=0$ and half-filling.
Reference system: decoupled set of Hubbard chains with 10 sites each.
The inset displays $\Omega(h)$ on a finer scale.
}
\label{1dchain}
\end{figure}

To test this, we have performed calculations for the one-dimensional Hubbard model.
The reference system consists of a decoupled set of finite Hubbard chains with 
$N_{\rm c}$ sites each.
Fig.~\ref{1dchain} shows the grand potential $\Omega$ as a function of the 
ficticious staggered field $h$ for $U=8$.
As one can see, the minimum of $\Omega$ is given by $h=0$, i.e.\ the V-CPT 
predicts the system to be a paramagnet, as expected physically.
We conclude that for this case quantum fluctuations are taken into account in
a sufficient way to prevent the system from becoming antiferromagnetic. 

\begin{figure}[t]
   \includegraphics[width=0.9\columnwidth]{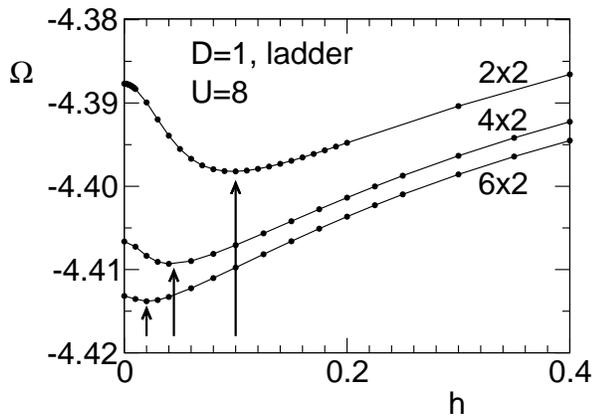}
\caption{$\Omega(h)$ as in Fig.\ \ref{1dchain} but for an infinite
one-dimensional Hubbard ladder. The reference system consists of 
finite ($4\times 2$ and $6\times 2$) ladders. 
Arrows indicate the optimal value of the ficticious staggered field.
}
\label{lad}    
\end{figure}

The results are not so straightforward if one considers a one-dimensional two-leg 
Hubbard ladder.
The reference system consists of decoupled finite ladders with $N_{\rm rung}$
rungs, i.e.\ $N_{\rm c} = N_{\rm rung} \times 2$.
Results for $N_{\rm rung} = 2$, $N_{\rm rung} = 4$ and $N_{\rm rung} = 6$ are 
shown in Fig.~\ref{lad}.
Despite the fact that the system is one dimensional, the calculations predict a 
finite value for the staggered field and for the staggered magnetization.
Clearly, this is an artifact of the remaining mean-field character on a longer
length scale.
However, we can see from Fig.~\ref{lad} that the optimal value of $h$ rapidly 
decreases when improving the approximation, i.e.\ with increasing size of the
clusters in the reference system.
This is consistent with the fact that no finite magnetization 
is expected in the $N_{\rm rung} \to \infty$ limit.

\begin{figure}[b]
    \includegraphics[width=0.9\columnwidth]{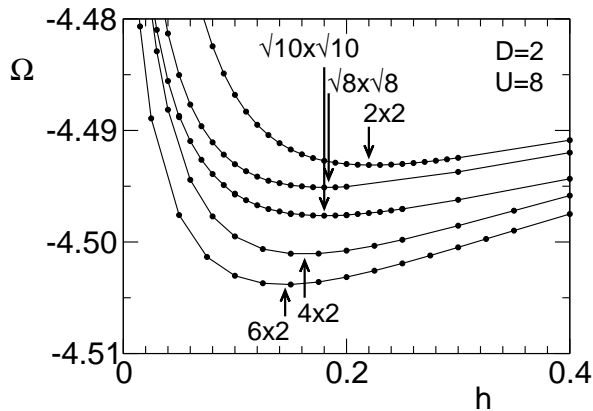}
\caption{$\Omega(h)$ as in Fig.\ \ref{lad} but now coupling the finite
ladders ($4\times 2$ and $6\times 2$) to a two-dimensional square 
lattice.
Results using the decomposition of the square lattice into $2 \times 2$, 
$\sqrt{8} \times \sqrt{8}$ and $\sqrt{10}\times\sqrt{10}$ clusters are 
shown for comparison.
Arrows indicate the optimal ficticious field.
}
\label{2d}    
\end{figure}

It is interesting to see what happens if one uses the $N_{\rm rung} \times 2$
ladders in order to build up a true {\em two}-dimensional system. 
The results are plotted in Fig.~\ref{2d}.
In this case, the optimal value of $h$ is much larger than in Fig.~\ref{lad}, 
and the order parameter (see below) remains finite and depends only weakly on 
the cluster size. 
This signals that for the two-dimensional system the antiferromagnetic state
is genuine, in contrast to $D=1$.

Fig.~\ref{2d} also shows the results for the two-dimensional lattice using 
different ``square'' clusters, $2 \times 2$, $\sqrt{8} \times \sqrt{8}$ and 
$\sqrt{10} \times \sqrt{10}$ (the $\sqrt{8} \times \sqrt{8}$ cluster is obtained
by discarding the rightmost sites in the first and the third line of the
$\sqrt{10} \times \sqrt{10}$ cluster shown in Fig.\ \ref{sqrt10}).
The comparison shows that convergence with respect to the function $\Omega(h)$ is
not yet achieved for the largest cluster size considered here.
In the limit of very large clusters the SFA becomes formally exact 
as the trial self-energy $\nv \Sigma$ is defined to be the exact self-energy of $H'$.
In this limit, we expect 
the location of the minima  ($\pm h_0$) of the function $\Omega(h)$ to go
to zero, or the function $\Omega(h)$ to become flat in a region around
$h=0$.
The reason is that in
the infinite system 
a finite value for the staggered magnetization will already be produced by an 
infinitesimally small field.
Note that for the series of $2 \times 2$, $4 \times 2$, $6 \times 2 , \dots$ 
clusters,
the reference system $H'$ does {\em not} approach the original two-dimensional
Hubbard model $H$.

\subsection{Order parameter}

\begin{figure}[b]
  \includegraphics[width=0.5\columnwidth]{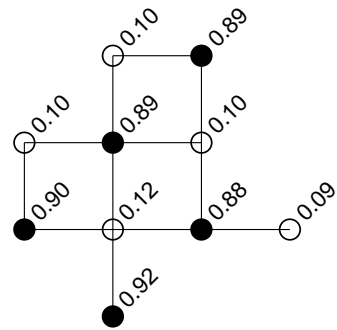}
\caption{
Local average occupation $\langle n_{\nv r, \uparrow} \rangle$ for $U=8$.
$\langle n_{\nv r, \downarrow} \rangle = 
1 - \langle n_{\nv r, \uparrow} \rangle$
(not plotted).
}
\label{nofi}
\end{figure}

While the staggered magnetization $m$ for the one-dimensional ladder system rapidly 
decreases with cluster size, $m$ remains finite and depends only weakly on the cluster 
size in case of the two-dimensional system. 
Differences in $m$ are found to be less than 1-2\% for the different cluster geometries 
considered in Fig.~\ref{2d}.
A relative difference $\Delta m / m \le 0.005$ is found when comparing the result
for the 10-site and the 8-site cluster.
The staggered magnetization is defined as 
$m = \partial \Omega / \partial h_{\rm ext}$ in the limit
$h_{\rm ext} \to 0$ where $h_{\rm ext}$ is the strength of an external 
{\em physical} staggered field (not to be confused with the ficticious field $h$).
Adding a respective field term to the Hamiltonian $H$ and performing the derivative 
with respect to $h_{\rm ext}$ of the grand potential (at the optimal ficticious 
field strength $h$), yields
$m = (1/N_{\rm c}) \sum_{\nv r} (-1)^{|\nv r|}
(\langle n_{\nv r,\up} \rangle - \langle n_{\nv r,\down} \rangle)$ where $\nv r$ 
runs over sites within a cluster, $N_{\rm c}$ is the number of cluster sites, and
$\langle n_{\nv r,\sigma} \rangle = (-1/\pi) \int_{-\infty}^0 d\omega \, 
\mbox{Im} \, G_{\nv r, \nv r, \sigma}(\omega+i0^+)$.
This is the usual expression for the staggered magnetization, but averaged over 
the cluster.
For the two-dimensional Hubbard model at $U=8$ we find $m \approx 0.80$.

Fig.\ \ref{nofi} shows the local average occupation $\langle n_{\nv r, \uparrow} \rangle$
for the sites $\nv r$ within the $\sqrt{10} \times \sqrt{10}$ cluster for $U=8$.
As any cluster approximation (constructed in real space), 
the V-CPT necessarily breaks the translational symmetries 
of the lattice:
The approximate self-energy is obtained from a translationally non-invariant reference
system which results from the decomposition of the original lattice into decoupled 
clusters of finite size.
This implies that the local Green's function, which is computed from the self-energy
using the Dyson equation, and thus the local occupations cannot be expected to be 
homogeneous (within a sublattice). 
It is interesting to see, however, that this is not a severe drawback:
Fig.\ \ref{nofi} shows that the variations of the spin-dependent local occupation and 
the local ordered moment are very moderate within a sublattice.

A much more inhomogeneous state with strongly varying local occupations is obtained
when coupling the ficticious field $h$ to two sites within the cluster only.
This variant has been considered using the $\sqrt{10} \times \sqrt{10}$ and the 
$\sqrt{8} \times \sqrt{8}$ clusters. 
In this case, too, a finite optimal value for $h$ and antiferromagnetic long-range
order are found (not shown).
The grand potential $\Omega$ at the optimal field, however, is considerably larger
than in the usual case where $h$ is coupled to all sites within a cluster.
This shows that despite the artificial breaking of translational symmetry, a 
homogeneous state is restored as far as possible.

\begin{figure}[t]
   \includegraphics[width=1\columnwidth]{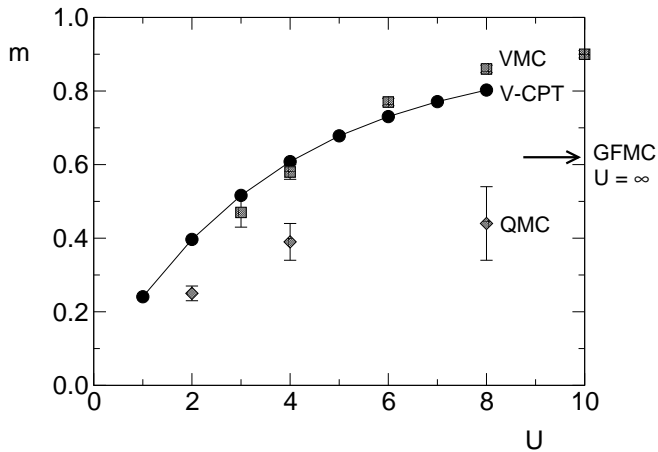}
\caption{
Comparison of the staggered magnetization $m$ as a function of $U$ at 
half-filling obtained by different methods:
variational CPT (V-CPT), variational Monte-Carlo (VMC) \cite{yo.sh.87b} 
and quantum Monte-Carlo (QMC), \cite{HT89} see text.
The arrow indicates the result $m=0.62 \pm 0.04$ of a Green's-function 
Monte-Carlo study \cite{TC89} for the two-dimensional Heisenberg model.
}
\label{smag}
\end{figure}

The $U$ dependence of the staggered magnetization is plotted in Fig.~\ref{smag} in 
comparison with the VMC results \cite{yo.sh.87b} (Gutzwiller-projected symmetry-broken 
trial wave function) and the results of auxiliary-field quantum Monte-Carlo (QMC). 
\cite{HT89}
Within the QMC, the order parameter is obtained from simulations of the static 
spin-spin correlation function at low temperatures.
It is assumed that the system effectively behaves as if at $T=0$ when
the thermal correlation length exceeds the cluster dimensions. \cite{HT89}
VMC and QMC data are extrapolated to $N_{\rm c} = \infty$. \cite{yo.sh.87b,HT89}

As one can see from Fig.~\ref{smag}, the variational CPT yields a staggered
magnetization which strongly disagrees with QMC data.
In the Heisenberg limit $U \to \infty$, the V-CPT seems to predict 
the staggered magnetization
to approach unity. However, whether it really becomes $m=1$
in this limit is not clear from our calculations yet.
On the other hand, physically, one would expect a reduction of the
staggered magnetization due to transverse spin fluctuations. In the
two-dimensional Heisenberg model,
several methods starting from the simplest spin-wave
theory up to Monte-Carlo methods, predict a reduction of the magnetization
to about $60\%
$ of its saturated value \cite{mano.91}.
On the other hand, the V-CPT agrees well with the results of the variational 
Monte-Carlo study, where transverse spin fluctuations are not fully taken into
account as well.
Furthermore, there is a very good qualitative agreement with respect to the size 
of $m$ and the trend of $m(U)$, when comparing with the results of a dynamical 
mean-field calculation \cite{ZPB02} (as the DMFT calculation has been performed
for the $D=\infty$ hypercubic lattice, one has to rescale the energies by a factor
four to obtain the same variance of the non-interacting density of states which
may serve as the energy unit).

This appears to be somewhat surprising since spatial correlations are neglected 
altogether in the DMFT and also in the VMC calculation where local Gutzwiller 
projectors are used, while the V-CPT does include the coupling to short-range 
correlations on the scale of the cluster size.
One has to bear in mind, however, that the size of the order parameter is strongly
affected by the coupling to {\em long}-range spin excitations.
Recall that in two dimensions and for any finite temperature the Mermin-Wagner 
theorem \cite{MW66,And52} shows that antiferromagnetic long-range order is destroyed 
due to spin waves with wave vector $\nv q \to 0$.
Hence, the overestimation of the staggered magnetization could be ascribed to 
the residual mean-field character of the V-CPT on a length scale exceeding the 
size of the cluster.
\footnote{
However, even
in this case one should expect some reduction of the
magnetization due to spin fluctuations
with a wavelength shorter than the cluster size.
}
This view is also substantiated by our results for the one-dimensional Hubbard 
ladder which have been discussed above:
To achieve a clear suppression of long-range order within the V-CPT, reference 
systems (finite ladders) as large as $6 \times 2$ have been required (Fig.\ \ref{lad}).
This is an indication that in two dimensions a $\sqrt{10} \times \sqrt{10}$ 
cluster might be to too small to include non-negligible effects of spin excitations 
on the order parameter.

There is another important point which has to be taken into account in this context:
For a cluster of a given size, an optimal V-CPT calculation should not only consider
the ficticious staggered field $h$ but {\em any} one-particle term in the Hamiltonian of
the reference system as a variational parameter.
It is in fact reasonable to assume that there is room for improvement: 
Consider, for example, the hopping between nearest neighbors within the cluster as
an additional variational parameter.
Actually, this has already been considered in Ref.\ \onlinecite{PAD03} for the $D=1$
Hubbard model.
There it was found that the optimal intra-cluster hopping is increased as compared to 
the nearest-neighbor hopping in the original lattice although the effect turned out 
to be rather weak.
Here, the situation is different due to the antiferromagnetic long-range order:
In the limit $U \to \infty$, an increased intra-cluster hopping implies 
an increased effective exchange interaction $|J|$.
Assuming the optimal ficticious field $h$ to be unchanged, this tends to 
{\em decrease} the order parameter $m$.
We have performed corresponding calculations which show that a variational adjustment 
of the intra-cluster hopping at a considerably increased value is very likely in fact.
However, a conclusive result has not yet been obtained. 
The reason is that the grand potential as a function of two variational parameters,
the ficticious staggered field and the intra-cluster hopping, becomes rather flat
in a wide region around the stationary point.
This requires an improved accuracy in the evaluation of the self-energy functional
which is difficult to achieve with a finite value for the (Lorentzian) broadening 
parameter $\delta$.
Work into this direction is in progress. \cite{pott.03u.se}

\subsection{Dynamical quantities}

While the V-CPT must be considered as mean-field-like on a length scale exceeding
the cluster size, it does account for short-range spatial correlations as the 
cluster problem is solved exactly.
For the two-dimensional Hubbard model at half-filling, short-range spin correlations 
are known to manifest themselves in dynamical quantities such as the local density 
of states.

\begin{figure}[t]
  \includegraphics[width=0.9\columnwidth]{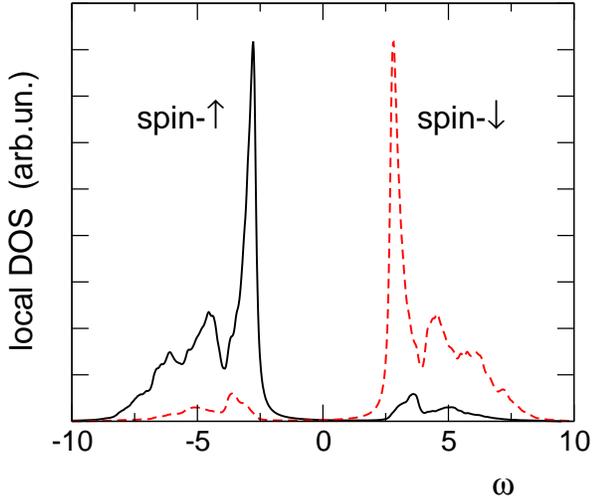}
\caption{
Spin-dependent
local density of states (DOS) $\rho_\sigma(\omega)$ 
(staggered average over the sites in a cluster) 
obtained for the same system as in Fig.~\ref{gse} via V-CPT
for $U=8$.
}
\label{dos}
\end{figure}

Fig.\ \ref{dos} shows the spin-dependent local density of states (DOS) 
$\rho_\sigma(\omega)$ for $U=8$ which is calculated as a staggered average over 
the sites in a cluster:
\beq
\rho_\sigma(\omega) = \frac{1}{N_{\rm c}} \sum_{\nv r} (-1)^{|\nv r|} \rho_{\nv r\sigma}
(\omega)
\eeq
where $\rho_{\nv r\sigma}(\omega) = (-1/\pi) \mbox{Im} \, G_{\nv r, \nv r, 
\sigma}(\omega+i0^+)$.
Roughly, the spectrum consists of two broad peaks around $\omega = \pm 5$ and 
two strong and narrow peaks at about $\omega = \pm 3$.
For both the high- and the low-energy excitations a strong spin polarization 
corresponding to $m \approx 0.80$ is clearly visible.
There is also some finite but low spectral weight within the insulating gap 
which, however, is an artifact of the finite Lorentzian broadening ($\delta = 0.1$).

The high-energy excitations in Fig.\ \ref{dos} are interpreted as charge excitations 
(Hubbard ``bands''). While these are due to local correlations,
the low-energy features (at $\omega = \pm 3$) result from 
(short-range) non-local correlations. 
The latter will be identified as being due to the coherent propagation of a quasiparticle, 
namely a ``spin bag''.
Physically, this spin bag originates from the frustration induced by the motion of 
the additional bare hole (electron) in the antiferromagnetic spin background. 
The different spectral features can easily be identified:
The high-energy features are due to the bare particle ``rattling around'' within the 
spin bag. 
This gives rise to an incoherent motion and broad energy ``bands'', i.e.\ the incoherent 
lower and upper 
Hubbard ``bands'' with a width set by the energy scale of the bare band width $W=8$.
The low-energy features, on the other hand, correspond to the above-mentioned coherent 
motion of the spin bag resulting in a strongly renormalized
quasi-particle band with a width essentially given by $2|J| = 8 t^2/U = 1$.

Note that these peaks are absent in a mean-field approach where off-site correlations 
are neglected altogether: 
A recent DMFT study \cite{ZPB02} of antiferromagnetic order shows a rather 
featureless DOS consisting of the two (polarized) Hubbard bands only.
Contrary, the effect of antiferromagnetic short-range correlations can be included 
in a cluster extension of the DMFT.
Additional structures appear in the DOS within the dynamical cluster approximation 
(DCA), for example.
Some indications of the mentioned low-energy features can be found by using the 
non-crossing approximation (NCA) to evaluate the DCA. \cite{MJPK00}
For a conclusive interpretation, however, the effects are too weak -- probably due 
to the limited cluster size (a $2 \times 2$ cluster in reciprocal space) and the 
finite temperatures considered.
 
\begin{figure}[t]
  \includegraphics[width=0.95\columnwidth]{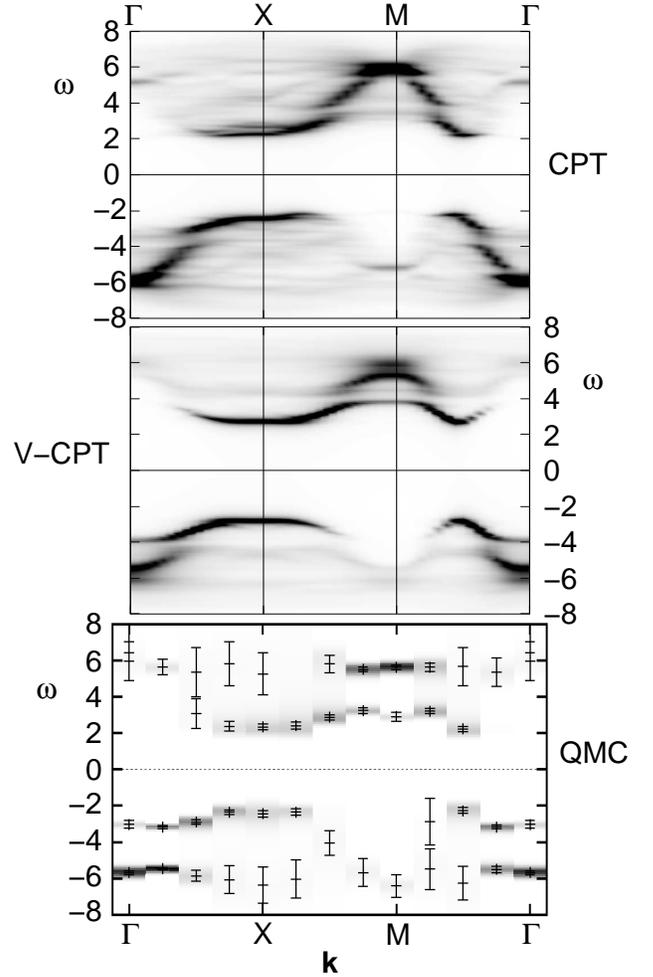}
\caption{
Density plot of the spectral function for the $D=2$ Hubbard model at $U=8$, 
half-filling and $T=0$ as obtained by the CPT with $h=0$ (top) and by the 
variational CPT with optimal ficticious staggered field $h \ne 0$ (middle).
The lattice is covered by $\sqrt{10} \times \sqrt{10}$ clusters.
Bottom: QMC (maximum entropy) result, taken from Ref.\ \onlinecite{GEH00}, for 
the same parameters but for a finite low temperature $T=0.1$ and an isolated 
$8\times 8$ cluster.
Dark (light) areas correspond to large (small) spectral weight.
}
\label{akw}
\end{figure}

More elucidating is a comparison of the $\nv k$-resolved spectral density with
available results from QMC simulations for isolated but larger clusters.
In order to illustrate this point, we have plotted in Fig.~\ref{akw} the 
spectral function $A_{\bm k}(\omega)$ for $U=8$ along high-symmetry directions 
in the Brillouin zone of the {\em chemical} lattice.
The result is compared with the result from the usual CPT ($h=0$) and with
numerically exact QMC data from Gr\"ober et al. \cite{GEH00} which are available
for an $N_{\rm c}=8\times 8$ isolated cluster and finite but low temperature 
($T=0.1$).
The spectral function $A_{\bm k}(\omega)$ obtained from the maximum-entropy method
(see Ref.\ \onlinecite{GEH00}) is shown in Fig.~\ref{akw} (bottom).
Since the spin-spin correlation length at $T=0.1$ considerably exceeds the 
cluster dimensions, the QMC result can be considered as a good approximation to 
the $T=0$ limit.
At half-filling the spectrum almost exactly respects the constraint 
$A_{{\bm k}}(\omega) = A_{{\bm k}+{\bm q}}(-\omega)$ with $\nv q = (\pi,\pi)$
which is predetermined by particle-hole symmetry.
This must be considered as a strong check of the numerics.
As for the finite system there is no spontaneous symmetry breaking, the spectrum
is spin independent and shows perfect translational symmetry with respect to the 
chemical lattice (periodic boundary conditions have been used).

This must be kept in mind when comparing with the V-CPT. 
In the V-CPT the real-space spectral function 
$A_{\nv R, \nv r, \nv R' \nv r', \sigma}(\omega) = -(1/\pi) \mbox{Im} \,
\langle \langle c_{\nv R, \nv r, \sigma} ; c^{\dag}_{\nv R', \nv r' \sigma} 
\rangle \rangle_\omega^{\rm (ret)}$ is spin dependent, and translational symmetry 
holds with respect to the superlattice vectors $\nv R$ only.
For a proper comparison with the QMC data we therefore compute
\beq
A_{\nv k}(\omega) = \frac{1}{L N_{\rm c}} \sum_{\nv R,\nv R'} \sum_{\nv r,\nv r'}
e^{i \nv k(\nv R + \nv r - \nv R' - \nv r')}
A_{\nv R, \nv r, \nv R' \nv r', \sigma}(\omega) \: ,
\eeq
see Fig.~\ref{akw} (middle).
If there was no antiferromagnetic order and no artificial breaking of translational 
symmetry due to the cluster approximation, i.e.\ if 
$A_{\nv R, \nv r, \nv R' \nv r', \sigma}(\omega)$ depended on the difference 
$\nv R + \nv r - \nv R' - \nv r'$ only, this would correspond to the usual
Fourier transformation. 
Here, $A_{{\bm k}}(\omega)$ is actually the diagonal element 
$A_{{\nv k, \nv k}}(\omega)$ obtained from two independent Fourier transformations
with respect to both, $\nv R + \nv r$ and $\nv R' + \nv r'$.
Taking the diagonal element, ensures a positive definite result, 
$A_{{\bm k}}(\omega) \ge 0$, and implies a spatial average over the cluster sites
(see Ref.\ \onlinecite{se.pe.02}).
Due to this spatial average, the spectral function is spin independent -- even
in the symmetry-broken state (an {\em integer} number of antiferromagnetic unit
cells are included in a single cluster).

The CPT spectral function is calculated accordingly but for $h=0$
(Fig~\ref{akw}, top).
This means that any signatures of long-range order are switched off in the 
spectrum, and only short-range correlations
(up to the cluster boundaries) are retained.
Both, the CPT and the V-CPT result, respect the condition
$A_{{\bm k}}(\omega) = A_{{\bm k}+{\bm q}}(-\omega)$ with $\nv q = (\pi,\pi)$
due to particle-hole symmetry.
Note that in both cases the spectral function is defined for any $\nv k$ point
in the Brillouin zone, contrary to the ``direct'' cluster method (QMC).

As already noted in the discussion of the local DOS, the V-CPT spectrum clearly
consists of four spectral features, two high-energy ``bands'' which show strong 
damping effects (incoherent Hubbard bands) and two narrow low-energy bands which 
represent the coherent dispersion of a quasi-particle (spin bag).
This four-band structure is also evident in the QMC result.
Comparing the energetic positions, dispersions, weights and widths of the four
spectral features, one can state that the agreement with the QMC spectrum is 
almost perfect.

Roughly, the CPT and the V-CPT spectra appear to be similar but looking at finer 
structures it is obvious that the CPT predicts a spectral function which is 
quite different:
First, and most important, there is no coherent low-energy band in the CPT 
spectrum.
This shows up when comparing with the V-CPT around $\Gamma$ for $\omega < 0$ 
(or around $M$ for $\omega > 0$), for example:
In agreement with the QMC result, the V-CPT predicts a dispersive low-energy band 
which extends continuously with spectral weight from $\Gamma$ to $X$ and which is
clearly separated from the more incoherent feature at higher energies.
On the other hand, in the CPT spectrum this is missing.
In the $\Gamma-M$ direction the low-energy features turn out to be too broad and 
are discontinuously split into several branches in the CPT spectrum.
The dispersion around $X$ is at variance with the QMC data.
Finally, at higher excitation energies, several weak and almost dispersionless 
bands can be found in the CPT spectrum while in the V-CPT there is a comparatively 
smooth incoherent background.
We conclude that the variational procedure is crucial to achieve a qualitatively 
correct reproduction of the one-particle excitation spectrum and of the coherent
quasi-particle band in particular.

The {\em physical} reason is as follows.
From previous QMC studies \cite{GEH00} it is well known that the quasi-particle 
band is the dispersion of a spin bag, i.e.\ an additional hole (electron) which is 
dressed by the local distortions of the spin order that are produced by the motion
of the hole in the antiferromagnetic background.
Since the linear extension of the spin bag is about 3-4 sites only,
this picture is already captured by an exact diagonalization of an isolated small 
cluster.
However, the emergence of a coherent band requires more, namely a coherent motion
of the spin bag on a larger length scale.
This is captured in the QMC results for a large cluster of $8 \times 8$ sites.
Of course, the perturbative treatment of the inter-cluster hopping within the 
CPT framework carries out a part of the job.
This results in a string dispersion in the V-CPT spectrum with a bandwidth of 
about $2|J| = 8 t^2/U = 1$ as can be read off from Fig.~\ref{akw}.
Also for the plain CPT the perturbative coupling of the clusters works into the
right direction:
Although the spectrum more or less consists of a two-band structure, there is 
a tendency towards the formation of a gap within each of the two bands, i.e.\
a coherent band {\em tends} to split off.
Within the plain CPT, however, the motion of the dressed hole cannot be completely
coherent as there is no definite alignment of spins across the cluster boundary.
Upon reaching the cluster edge, the spin bag encounters a misaligned spin with 
50\% probability and is partly reflected back inside the cluster.
This partial loss of coherence explains the several bands at higher energies in the 
CPT spectrum which are absent in the V-CPT.
The variational generalization of the CPT cures this problem by ordering spins 
antiferromagnetically with help of the ficticious staggered field not only within 
but also across the cluster boundaries thereby allowing the coherent spin-bag 
propagation.

\section{Conclusions and outlook}
\label{conc}

Correlated electron systems in two dimensions are in many cases characterized
by strong local and non-local but still short-ranged correlations on the one hand 
and by long-range, e.g.\ magnetic, order on the other.
Here, we have presented a variational extension of the cluster-perturbation 
theory which combines the exact diagonalization of isolated small clusters with 
a mean-field concept to build up an infinite lattice.
Conceptually, the method is based on the recently proposed self-energy-functional
approach (SFA) which sets up a very general variational scheme to use dynamical 
information from an exactly solvable reference system $H'$ (the isolated cluster)
to approximate the physics of a system $H$ (the $D=2$ Hubbard model) in the
thermodynamic limit.

We have applied the variational CPT (V-CPT) to the Hubbard model at half-filling
to study the antiferromagnetic phase at zero temperature.
The diagonalization of Hubbard clusters of finite size (typically $N_{\rm c}=10$) 
is performed using the standard Lanczos algorithm.
In comparison with results from variational Monte-Carlo and quantum Monte-Carlo
studies, the V-CPT predicts the ground-state energy and related static quantities 
with high accuracy.
While long-range antiferromagnetic order is obtained for the $D=2$ model, the
V-CPT yields a paramagnetic state for $D=1$.
This indicates that quantum spin fluctuations which inhibit an ordered phase
in the $D=1$ case are included properly.
For one-dimensional Hubbard ladders the method in principle
incorrectly predicts antiferromagnetic order;
however, the staggered magnetization is small and tends to vanish when increasing 
the number of rungs in the cluster (up to $2 \times 6$).
The finite but small $m$ for the ladder system should be considered as a mild 
reminiscence of a typical mean-field artifact which shows up because longer-range
spin correlations exceeding the cluster dimensions are neglected.
A similar effect is seen for the $D=2$ system:
Here the approximation is even stronger because the linear dimension of the cluster 
must be reduced even more to keep the number of cluster sites $N_{\rm c}$ 
constant.
For the $D=2$ system, antiferromagnetic order is expected physically and is also 
found by the calculations.
However, longer-ranged spin correlations give rise to a considerable reduction 
of the order parameter which is not seen in the V-CPT for the maximum cluster 
size that has been considered.

An important advantage of our method is that local and off-site short-range 
correlations within the ordered phase can be treated exactly.
This shows up when looking at dynamical quantities, such as the spin-dependent 
local density of states or the spectral function $A_{\nv k}(\omega)$.
The spectral function as calculated from our self-consistent cluster approach 
agrees extremely well with the QMC (maximum-entropy) result for an $8 \times 8$
Hubbard lattice at finite but low temperatures.
In particular, it is possible to reproduce the dispersions, widths and weights
of the different spectral features.
This is due to the fact that the typical four-band structure arises not only from local
correlations which are captured in dynamical mean-field theory, for example, but
also from a strong coupling to off-site spin correlations.
The formation of a spin-bag quasi-particle as a hole which is dressed by the 
distortions of the antiferromagnetic spin structure that are introduced 
due its motion, is contained in the exact treatment of the cluster.
In addition, the mean-field coupling of the individual clusters mediates the
information on the spin order across the cluster boundaries and thereby gives
a qualitatively correct description of the coherent propagation of the quasiparticle.
This is essential to reproduce the low-energy quasi-particle band in the spectrum.

There are different interesting routes to be explored in the future:
A straightforward but technically ambitious idea is to employ a quantum Monte-Carlo
technique within the self-energy-functional framework. 
This would offer the possibility to use a decomposition of the lattice into larger
clusters which is expected to be important for a reliable estimate of the order
parameter for example.
Another straightforward extension of the method concerns the number of variational
parameters.
For our present purposes the consideration of a single parameter, the ficticious
staggered field $h$, has been sufficient. 
It is obvious, however, that the results should improve when treating {\em any} 
one-particle parameter within the cluster as a variational parameter.
This might also affect the magnetic properties because of the direct link between 
hopping parameters and the effective exchange interactions.
Here an improved numerical evaluation of the theory is required which is free from 
the use of broadening parameters, see Refs.\ \onlinecite{Pot03a,pott.03u.se}.

An important conceptual problem of the method consists in the fact that a suitable
reference system can be found in case of local (on-site) interaction terms only.
The reason for this restriction is that for models including e.g.\ a nearest-neighbor 
Coulomb interaction, the partitioning of the infinite lattice into decoupled clusters 
is impossible without cutting the inter-cluster interaction. 
General arguments, \cite{Pot03a} however, require that the interaction part of the 
Hamiltonian of the reference system be unchanged.  
Consequently, there is a need for an extension of the theory to models with 
non-local interaction terms. 
This will be the subject of a forthcoming paper. \cite{AEvdLP04}

The V-CPT compares with the recent cluster extensions of the DMFT:
\cite{SI95,HTZ+98,LK00,ko.sa.01,ma.ja.02,bi.ko.02}
Characteristic to both approaches is the combination of a numerically exact 
treatment of an isolated cluster with an approximate mean-field treatment of 
the coupling between different clusters.
The V-CPT is, however, much easier to implement numerically and in principle
allows for a
diagonalization of larger clusters which facilitates a proper finite-size scaling.
The reason for this conceptual simplicity is the fact that there is no coupling
to bath degrees of freedom which is inherent to the cluster extensions of the DMFT.
Recent results \cite{PAD03} for the $D=1$ Hubbard model have shown that it is in
fact more efficient to consider larger clusters instead of a coupling to bath sites. 
For the two-dimensional case, however, there is no answer to this question at 
present, and it is unclear whether or not bath degrees of freedom efficiently 
speed up the convergence to the exact solution with increasing cluster size.
Note that a hopping to uncorrelated bath sites is nothing but a modified 
one-particle part of the Hamiltonian of the reference system and thus completely 
in line with the general concept of the self-energy-functional approach.
In particular, it can be very interesting to consider an extension of the V-CPT
with a coupling to a {\em few} bath sites as has been done in previous studies.
\cite{Pot03a,PAD03}
There are at least two reasons for that:
First, this bridges the gap to the cellular dynamical mean-field theory (C-DMFT) 
\cite{ko.sa.01} which is obtained in the limit of infinite number of bath degrees 
of freedom.
Second, bath sites are expected to become important for the study of doped systems
as they can serve as a particle reservoir.

For doped systems, the variational optimization of one-particle parameters should
include the on-site energy of the cluster sites as well as the additional consideration 
of a few bath sites.
This is necessary to realize fillings which are not commensurate with the cluster 
size and to achieve smooth doping dependencies in the entire doping range.

In the present paper we have focused on a ficticious staggered magnetic field
as a variational parameter to describe antiferromagnetic long-range order.
Moreover, one can envisage to introduce in the same way a ficticious pairing field  
in order to study off-diagonal long-range order and superconductivity.
While this introduces the numerical difficulty of diagonalizing clusters without
fixing the particle number, such an approach offers the exciting perspective of 
analyzing the effects of short-range correlations in the superconducting phase.

\acknowledgments
This work is supported by KONWIHR (CUHE), by the Doctoral Scholarship 
Program of the Austrian Academy of Sciences (M.A.), by the Deutsche 
Forschungsgemeinschaft via Heisenberg grant AR 324/3-1 (E.A.),
via project HA 1537/20-1 and via the Sonderforschungsbereich 410.

\end{document}